\documentclass[11pt]{article}

\usepackage[utf8]{inputenc}
\usepackage[T1]{fontenc}
\usepackage{mathptmx}
\usepackage[scaled=0.90]{helvet}

\usepackage[a4paper,top=2.3cm,bottom=2.3cm,left=2.2cm,right=2.2cm]{geometry}
\usepackage{microtype}
\usepackage{graphicx}
\usepackage{booktabs}
\usepackage{array}
\usepackage{tabularx}
\usepackage{multirow}
\usepackage{caption}
\usepackage[table]{xcolor}
\usepackage{enumitem}
\usepackage{setspace}
\usepackage{authblk}
\usepackage{titlesec}
\usepackage{siunitx}
\DeclareSIUnit{\angstrom}{\AA}
\usepackage{textcomp}
\usepackage[numbers,sort&compress,super]{natbib}
\usepackage[colorlinks=true,allcolors=blue!55!black]{hyperref}

\definecolor{naturetext}{HTML}{1F2933}
\definecolor{accent}{HTML}{1F2933}
\definecolor{docnotelinkcolor}{HTML}{2B4A8B}

\titleformat{\section}{\large\sffamily\bfseries\color{accent}}{\thesection}{0.6em}{}
\titleformat{\subsection}{\normalsize\sffamily\bfseries\color{naturetext}}{\thesubsection}{0.5em}{}
\titleformat{\subsubsection}{\small\sffamily\bfseries\color{naturetext}}{\thesubsubsection}{0.5em}{}
\titlespacing*{\section}{0pt}{1.0ex plus 0.3ex minus 0.2ex}{0.5ex}
\titlespacing*{\subsection}{0pt}{0.9ex plus 0.3ex}{0.4ex}
\titlespacing*{\subsubsection}{0pt}{0.7ex plus 0.3ex}{0.35ex}

\newcolumntype{R}{>{\raggedleft\arraybackslash}X}
\newcolumntype{C}{>{\centering\arraybackslash}X}
\title{\sffamily\bfseries\boldmath G2P Explorer: A Native iOS Framework for
Residue-Level Genomics to Proteomics Visualization and Structural Variant
Interpretation}

\author[1]{Arifa Akter Eva}
\author[1]{Md Abrar Hamim}
\author[1,$\ast$]{Md. Manzurul Hasan}
\affil[1]{\normalfont\small Department of Computer Science, American
International University-Bangladesh, Dhaka, Bangladesh}
\affil[$\ast$]{\normalfont\small Corresponding author:
\href{mailto:manzurul@aiub.edu}{manzurul@aiub.edu}}

\date{}

\begin{document}
\maketitle
\vspace{ 2.2em}
\noindent\rule{\linewidth}{0.4pt}
\vspace{0.2em}

\noindent\textbf{\sffamily Abstract}\\[0.3em]
\noindent
Genetic testing reports coding variants far faster than they can be interpreted,
and placing a variant in its biophysical context, the domain it perturbs, whether
its residue is buried or exposed, whether it lies near a disulfide bond or a
predicted binding pocket increasingly requires projecting it onto a three dimensional protein model. The Genomics 2 Proteins (G2P) portal unifies the gene-to-structure identifier chain with a dense, residue indexed annotation table,
but its visualization layer presumes a desktop browser and is awkward at the
bedside, in the classroom, or in the field, where a phone or tablet is often the
only device. We present G2P Explorer, a native iOS framework that consumes
the public G2P REST API on device, parses its seventy one column tab separated
feature tables without loss of fidelity, and presents the result through six
interlinked modules sharing a single observable view model. Beyond porting, it
contributes a SwiftUI Canvas multi-track sequence renderer, an on device
reconstruction of the portal's unavailable isoform alignment route, a bidirectional
Swift  JavaScript structural bridge that absorbs the AlphaFold model file versioning
scheme, and a fault tolerant ingestion layer that parses semi structured free text
and distinguishes absent annotations from zero. Each searched protein is cached on
device after the first fetch, so it reopens instantly and works offline, and the
embedded structural view is drawn in a reduced form suited to a small screen. Across
six proteins spanning 189-1{,}863 residues, the framework sustains interactive
frame times (3.1-16.6\,ms) and modest memory (16-72\,MB). G2P Explorer is an open,
reproducible mobile companion to the G2P portal for hypothesis generation, teaching,
and on the go variant interpretation.

\vspace{0.4em}

\section{Introduction}

Contemporary human genetics is shaped by a widening gap between the rate at which
genetic variation is catalogued and the rate at which it can be functionally
understood. Population-scale sequencing, clinical screening, and deep mutational
scanning now describe millions of coding variants, yet only a small fraction carry a
confident functional or clinical
interpretation \citep{McInnes2021,Landrum2018,Chen2024}. The limiting step is rarely
the size of the catalog but the difficulty of situating a variant in its
structural and biophysical neighborhood. Whether a substitution destabilizes a
fold, abolishes a catalytic contact, exposes a buried core position, or disrupts a
splice-dependent isoform are structural questions posed on top of genomic
coordinates, and answering them requires mapping the variant onto a
three-dimensional model of the affected protein \citep{Cotto2023,Cheng2023}.

The release of accurate predicted structures has largely closed the
structural coverage half of this problem. Where the Protein Data
Bank\citep{Berman2000} historically offered experimentally solved coordinates for
only part of the human proteome, AlphaFold and related systems now provide
per-residue confidence-scored models for essentially every reviewed
sequence\citep{Jumper2021,Lin2023,Abramson2024,Krokidis2025}, distributed at scale
through the AlphaFold Protein Structure Database\citep{Varadi2024}, and
structure-aware variant effect predictors such as AlphaMissense build directly on
this foundation\citep{Cheng2023}. The bottleneck has thus shifted from
whether a structure exists to how fluently a researcher can travel
from a genomic coordinate to a structural residue while preserving the chain of
intermediate identifiers: Ensembl gene and transcript, MANE Select
reference \citep{Morales2022}, RefSeq accession, UniProt entry \citep{UniProt2023},
isoform, and finally a PDB or AlphaFold model.

The Genomics 2 Proteins (G2P) portal was designed to remove precisely this
friction \citep{Kwon2024}. For any human gene, it assembles a table that joins the
full identifier chain and pairs it with a residue indexed annotation table
aggregating secondary structure \citep{Kabsch1983}, AlphaFold per residue confidence (pLDDT), accessible surface area (ASA), backbone dihedral angles, UniProt domain and feature lanes, post translational modifications, predicted druggable pockets from fpocket and p2rank \citep{LeGuilloux2009,Krivak2018}, and intra  and inter chain interaction networks. The portal surfaces these products through interactive sequence tracks and an embedded Mol* viewer \citep{Sehnal2021}, and exposes them programmatically through a public REST interface that returns tab-separated
tables.

Although the desktop experience is feature complete, three classes of users are
poorly served by a browser only interface. Clinicians and genetic counsellors
increasingly consult variant evidence on a phone or tablet at the point of
interaction; students and trainees work in classrooms where driving a desktop
browser is cumbersome; and researchers in journal clubs, conferences, or field
settings need to inspect a candidate variant's structural neighbourhood during a
conversation, without opening a laptop. For all three, a mobile native experience is
not a degraded desktop tool but a distinct interaction modality with its own
ergonomics. Established environments
(PyMOL \citep{Yuan2017}, ChimeraX \citep{Pettersen2021}, Mol* \citep{Sehnal2021},
ProtVista \citep{Watkins2017}, IGV \citep{Thorvaldsdottir2013}) were optimised for
large monitors, pointer input, and abundant memory; their grammar of
drag to rotate, hover to tooltip, and dense multi pane layouts translates poorly to
touch surfaces and small viewports. A genuinely mobile native framework must
therefore rethink the rendering stack, the interaction grammar, and the information
architecture rather than re skin the desktop interface.

Re implementing an aggregating scientific web service as a mobile native instrument
proved not to be a matter of transcription. Three engineering obstacles were
decisive, and we foreground them because they recur whenever a rich web service is
brought to the phone. First, the portal's isoform alignment endpoint returns
HTTP~404 to programmatic clients the alignment a user sees on the desktop is in
fact computed in the browser from UniProt FASTA sequences, so a faithful mobile
client must carry its own pairwise aligner. Second, linking a touch-native
two-dimensional sequence view to a three-dimensional structure rendered inside a web
view crosses a language and process boundary in both directions, and the AlphaFold
model file versioning scheme advanced during development, so a hard-coded structure
URL silently breaks. Third, a substantial fraction of the portal's richest
annotation (disulfide partners, pocket geometry, and per-isoform coverage) arrives as
free text or is simply absent for noncanonical isoforms, so faithful rendering
requires parsing semi-structured strings and visibly distinguishing absence from
zero. We treat these as first-class design problems and revisit each as a worked
case study.

\begin{enumerate}
    \renewcommand{\labelenumi}{\roman{enumi}.}

\item \textbf{A mobile native interaction grammar for genomics to proteomics
exploration}, organised as six modules (search, sequence, structure, variant
landscape, isoform mapping, druggable pockets) that share a single
\texttt{@Observable} view model and support deep cross navigation, so a residue
selected in one module is highlighted in all others. To our knowledge this is the
first iOS native client of the G2P portal.

\item \textbf{A SwiftUI \texttt{Canvas} multi-track sequence renderer} that draws up
to six meaningful feature tracks for proteins of up to $\sim$2{,}000 residues at
sustained interactive frame rates, using a per-residue immediate-mode draw strategy
chosen over view stacking and single-path alternatives.

\item \textbf{A client-side isoform alignment pipeline} compensating for an
unavailable API route: the app fetches canonical and alternative isoform FASTA
sequences from UniProt in parallel, computes a Needleman--Wunsch alignment on
device\citep{Needleman1970}, and re-anchors the result to canonical coordinates so
indexing stays stable across modules.

\item \textbf{A bidirectional Swift--JavaScript structural bridge} in which a residue
tapped in the sequence view is focused and marked in the 3D viewer, pLDDT confidence
bands are filtered by partitioning model atoms on their B-factor, and the evolving
AlphaFold model file versioning is absorbed by a cascading fallback loader.

\item \textbf{A fault-tolerant seventy-one-column ingestion layer} that preserves
non-ASCII and punctuated headers verbatim, distinguishes an absent annotation from a
value of zero, and parses semi-structured free-text cells into typed records,
omitting rather than fabricating data when a cell fails to parse.

\item \textbf{An interactive analytical layer}: a pLDDT versus ASA quadrant scatter,
a mobile-native Ramachandran panel for arbitrary proteins, and an honest per-isoform
coverage view, with a quantitative benchmark across six proteins and an offline
demonstration routed through the identical parsing and rendering path as live
responses.

\item \textbf{A persistent offline cache and a reduced structural view for small
screens.} After a protein is fetched once, its parsed feature and isoform data are
written to on-device storage, so it reopens from local storage with no further
request and stays usable offline, extending offline use beyond the single bundled
sample to every protein the user has searched. The embedded structural view also
adopts a reduced rendering that keeps only the elements needed to read a residue in
context, because the dense desktop layout is hard to interpret on a small screen.

\end{enumerate}

\section{Results}
This section presents the experimental evaluation and demonstrates the effectiveness of the framework through a series of quantitative and qualitative analyses.

\subsection{An integrated mobile resource and its workflow}
\begin{figure}[!tb]
\centering
\includegraphics[width=0.74\linewidth]{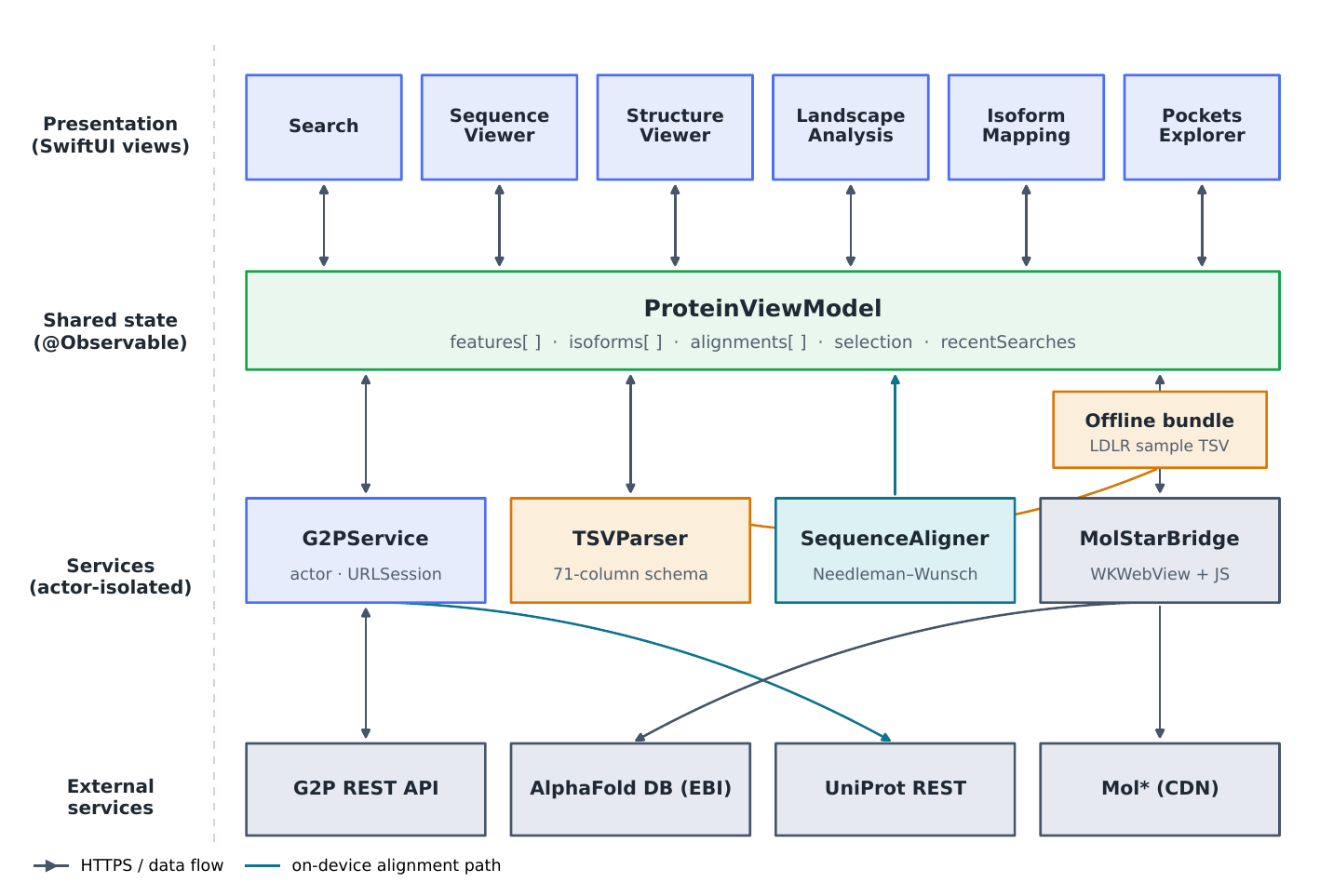}
\caption{\textbf{G2P Explorer system architecture.} Six SwiftUI modules (Search,
Sequence Viewer, Structure Viewer, Landscape Analysis, Isoform Mapping, Pockets
Explorer) share a single \texttt{@Observable} \texttt{ProteinViewModel} that owns
the parsed feature, isoform, alignment, and selection state. The actor isolated
\texttt{G2PService} issues concurrent HTTPS requests to the G2P REST API, the EBI
AlphaFold service, and the UniProt REST service; the content agnostic
\texttt{TSVParser} handles every tab separated endpoint; the on device
\texttt{SequenceAligner} reconstructs pairwise isoform alignments that the portal
does not serve to programmatic clients; and the \texttt{MolStarBridge} drives a
Mol* viewer hosted inside a \texttt{WKWebView}. A bundled LDLR sample and an
on device cache of previously searched proteins support identical behaviour offline
for the four non structural modules.}
\label{fig:arch}
\end{figure}

G2P Explorer exposes the residue-level data products of the G2P portal as an
integrated, touch-driven workflow on iOS (Fig.~\ref{fig:arch}). A session begins in
the Search module, where the user enters a gene symbol or UniProt accession or taps
one of eight curated preset chips spanning a range of lengths and disease contexts
(KRAS, TP53, LDLR, DNMT3A, MORC2, BRCA1, EGFR, BRCA2). On selection, the shared view
model dispatches concurrent requests to the G2P endpoints (protein features and the
gene  transcript  protein  isoform  structure map), streams each response through the
seventy-one-column parser, and publishes typed Swift records; the remaining five
modules read the same state and update declaratively. One qualification, examined in
Section~\ref{sec:cs1}: the isoform alignment request is not a single network call
because that route is unavailable; it is realised as an on-device reconstruction,
whereas the feature and mapping requests remain genuine fetches.

This design positions G2P Explorer as a mobile native \emph{resource} rather than
a single purpose application: each module is a re thought visualization idiom, and
the system together supports a hypothesis generation loop that, on the desktop,
would normally span several browser tabs. Selecting a residue in the Sequence
module focuses the same residue in the Structure module and highlights it in the
Landscape, Isoform, and Pockets modules; conversely, tapping a pocket card scrolls
the Sequence module to the pocket residues and overlays a spacefill representation
in the Mol* viewer.

\subsection{Native iOS architecture and asynchronous data flow}
The framework is implemented in Swift~5.9 with SwiftUI for iOS~16 and above,
organised into four layers (Fig.~\ref{fig:arch}): six SwiftUI views; a shared state
layer around a single \texttt{@Observable} view model; a services layer with an
actor isolated networking client (\texttt{G2PService}), a content agnostic parser
(\texttt{TSVParser}), an in process aligner (\texttt{SequenceAligner}), and a Mol*
bridge (\texttt{MolStarBridge}); and an external layer of the G2P, AlphaFold, and
UniProt services plus a bundled offline sample. The only third party runtime
dependency is the Mol* JavaScript bundle, loaded into a \texttt{WKWebView} from a
CDN; the Swift code itself uses no third party packages.

\begin{figure}[!tb]
\centering
\includegraphics[width=0.74\linewidth]{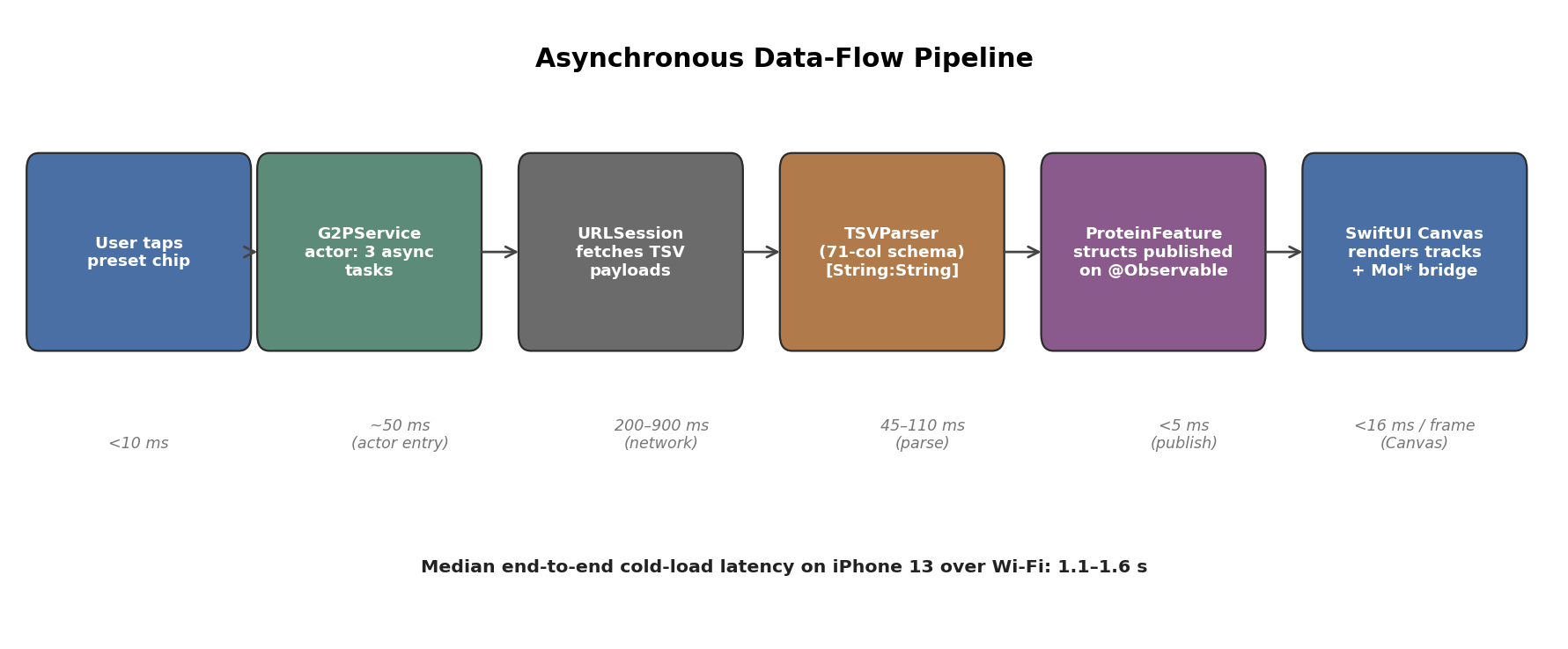}
\caption{\textbf{Asynchronous data flow pipeline.} A gesture in the Search module
triggers concurrent fetches through the \texttt{G2PService} actor; responses are
parsed by the \texttt{TSVParser}, lifted into typed Swift structs, published on
the \texttt{@Observable} view model, and rendered by the SwiftUI \texttt{Canvas}
and the Mol* bridge. Per stage latencies were measured on an iPhone~13 over Wi Fi
for the LDLR payload; median end to end cold load was 1.1 1.6\,s, dominated by
the network stage rather than by parsing or rendering.}
\label{fig:pipeline}
\end{figure}

\begin{figure}[!tb]
\centering
\includegraphics[width=0.62\linewidth]{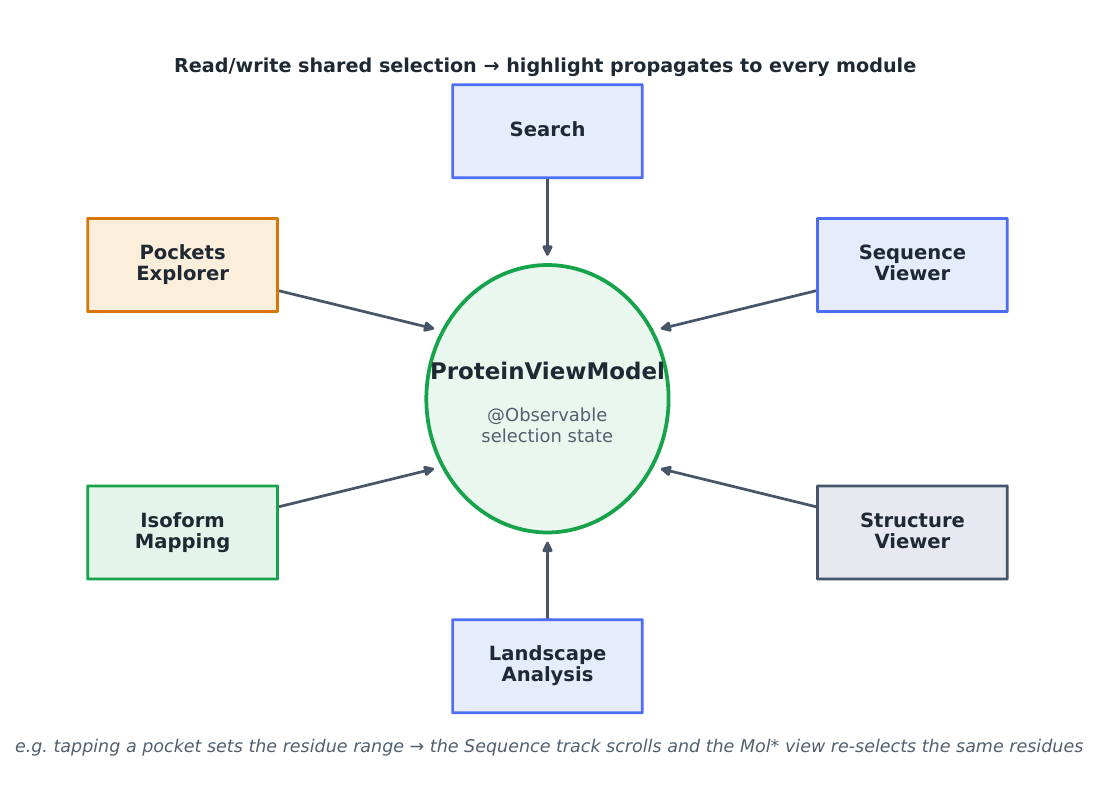}
\caption{\textbf{Cross module residue linking through the shared view model.}
Every module both reads and writes the selection state on
\texttt{ProteinViewModel}, so a residue or pocket selected in one module appears
highlighted in all others without explicit message passing or data re fetching.
Tapping a pocket card sets the residue range, scrolls the Sequence module, and
re issues a Mol* selection in the Structure module; tapping a residue in the
Sequence module opens a detail sheet listing every populated column for that
position (Supplementary Fig.~1).}
\label{fig:cross}
\end{figure}

The six modules are hosted in a \texttt{TabView} bound to one view model instance.
The \texttt{@Observable} macro provides fine grained dependency tracking, so a view
recomputes only when a property it reads is mutated. The view model owns the current
gene and accession, the parsed feature, isoform, and alignment arrays, the selected
residue or range, and a small recent search history in \texttt{UserDefaults}. Beyond
this history, parsed feature and isoform data are written to an on device store after
the first successful fetch, keyed by gene symbol or accession, so a later visit reads
from local storage and works offline for the four non structural modules.
Cross module navigation reduces to writing one or two properties that the destination
module reads declaratively (Fig.~\ref{fig:cross}). Network traffic is concentrated in
the \texttt{G2PService} actor (Fig.~\ref{fig:pipeline}); actor isolation prevents
concurrent requests from corrupting \texttt{URLSession} state, and
structured concurrency cancellation propagates from the SwiftUI lifecycle, so
switching tabs during a load cancels the in flight task. Errors surface as a single
typed enum with localised descriptions shown inline.

\subsection{Multi track sequence visualization on a touch surface}
\begin{figure}[!tb]
\centering
\includegraphics[width=0.76\linewidth]{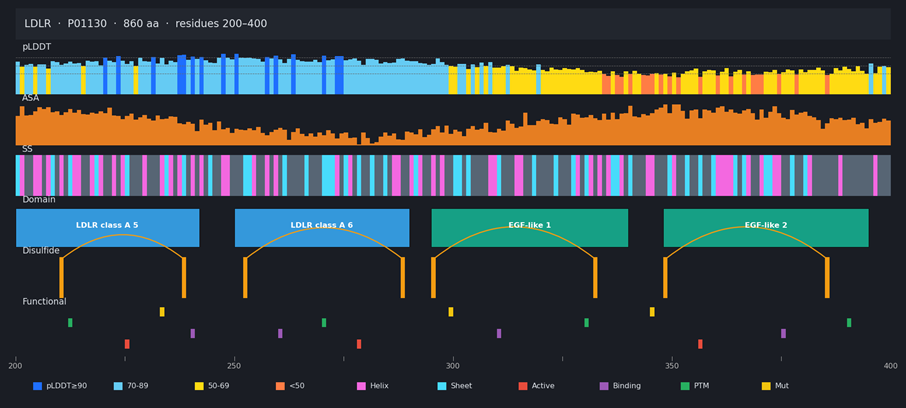}
\caption{\textbf{Multi track sequence renderer for LDLR residues 200 400.} Six
stacked tracks share a common residue axis: pLDDT (coloured by the AlphaFold
confidence palette, with dashed reference lines at 50/70/90), accessible surface
area normalised to the dataset maximum, three state secondary structure, merged
UniProt domain features, disulfide bonds rendered as on canvas B\'ezier arcs
between bonded cysteines, and a four lane functional site track (active site,
binding site, post translational modification, mutagenesis). The view supports
pinch zoom (one  to eight times), momentum scroll, and tap to detail; residues are
drawn at one filled rectangle per residue per track without binning.}
\label{fig:sequence}
\end{figure}

The Sequence module is the centrepiece of the framework. It renders six stacked
annotation tracks against a shared residue axis (Fig.~\ref{fig:sequence}): a
per residue pLDDT bar chart, an accessible surface area track, a three state
secondary structure track, a merged UniProt domain lane, a disulfide bond track,
and a four lane functional site track. All residues remain in the zoomable range;
no binning or paging is applied. Each track is drawn in immediate mode through
SwiftUI's \texttt{Canvas} primitive. During development we rejected two naive
alternatives: a tall stack of \texttt{Rectangle} views, which incurs full layout
cost per residue and drops the frame rate at roughly eight hundred residues across
six tracks, and a single accumulated \texttt{Path} per track, which forfeits
per residue colour expressivity. The chosen design issues one filled rectangle per
residue per track through \texttt{context.fill(Path(rect))}, with colour resolved
by a compact lookup table; the \texttt{Canvas} body re executes on zoom and
scroll, but the work per frame is bounded by the number of residues in the visible
viewport.

The pLDDT track follows the AlphaFold palette (very high blue $\geq$90, confident
green 70  89, low amber 50  69, very low red $<$50) with bar heights proportional to
confidence, so both the overall distribution and local dips are visible at a glance.
The ASA track shares this encoding but normalises to the dataset maximum
($\sim$\SI{260}{\angstrom\squared} for LDLR), keeping scale comparable across
proteins. Five UniProt feature categories (Domain, Repeat, Signal, Transmembrane,
Topological domain) are merged into one lane, because vertical space is scarce and
these rarely co occur on a residue; contiguous runs are drawn as rounded rectangles,
labelled in place when wider than sixty points. Disulfide bonds arrive as free text
(e.g.\ ``Disulfide bond 27 39''); a regular expression extracts the partner and a
quadratic B\'ezier arc renders the bond on the same canvas, exposing short loops and
long range bonds directly. To our knowledge no other mobile sequence viewer surfaces
disulfide topology natively.

\subsection{Structural and landscape visualization}
\begin{figure}[!tb]
\centering
\includegraphics[width=0.68\linewidth]{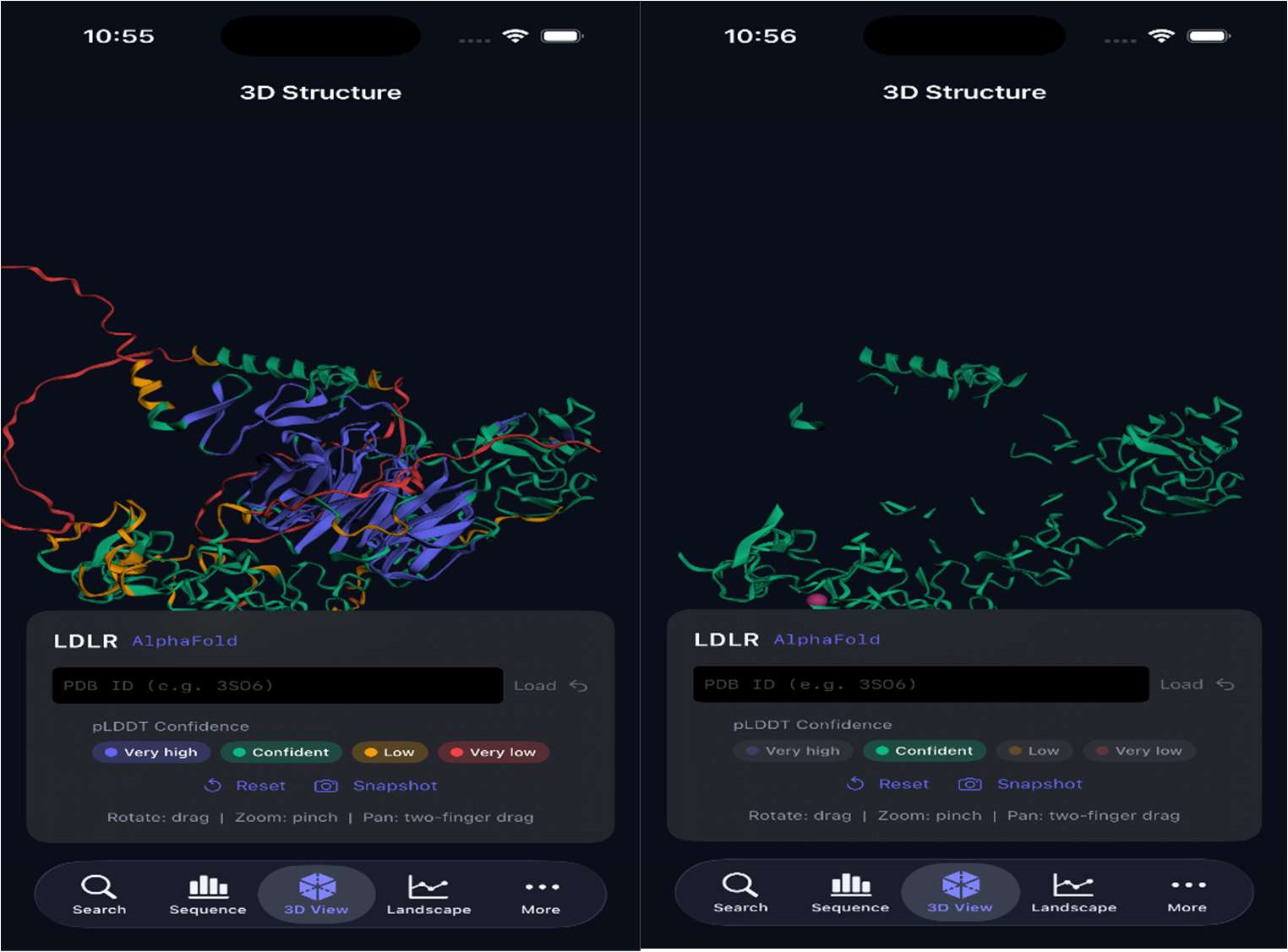}
\caption{\textbf{Embedded Mol* viewer driven from Swift.} The Structure module
hosts Mol*~4.9.0 inside a \texttt{WKWebView}; the Swift layer drives the viewer
through \texttt{evaluateJavaScript} and subscribes to \texttt{molLoaded} messages
via a \texttt{WKScriptMessageHandler}. AlphaFold mmCIF files are fetched from the
EBI AlphaFold service. Left: LDLR rendered with pLDDT confidence colouring (very
high to very low). Right: a single residue brought into spatial focus and marked
from a tap in the Sequence module. Confidence band filtering is realised by
partitioning model atoms on their B factor field
(Section~\ref{sec:cs2}).}
\label{fig:structure}
\end{figure}

Three dimensional rendering is delegated to Mol*\citep{Sehnal2021} (chosen over the
lighter 3Dmol.js\citep{Rego2015} for its richer representation set and its role as
the de facto AlphaFold and PDB web viewer), embedded in a \texttt{WKWebView} in the
Structure module (Fig.~\ref{fig:structure}). A bundled HTML page imports Mol*~4.9.0
from a CDN, initialises the viewer in a full screen container, and exposes a minimal
JavaScript surface to Swift. Because the HTML shell is read from the app bundle, the
viewer initialises offline; only the mmCIF download and the library require network.
A control sheet exposes pLDDT, chain, and rainbow colouring, functional site and
disulfide toggles, and a reset. Because the embedded viewer inherits a desktop layout
that becomes crowded on a phone, the structural view shows a reduced set of elements:
the cartoon model, the active colouring, and any focused residue, while omitting the
secondary panels and overlays that add clutter without aiding interpretation. Its
biological value is to situate a residue observed in the Sequence module within its
spatial context, where surprises emerge: residues distant in sequence may be adjacent
in the fold.

\begin{figure}[!tb]
\centering
\includegraphics[width=0.60\linewidth]{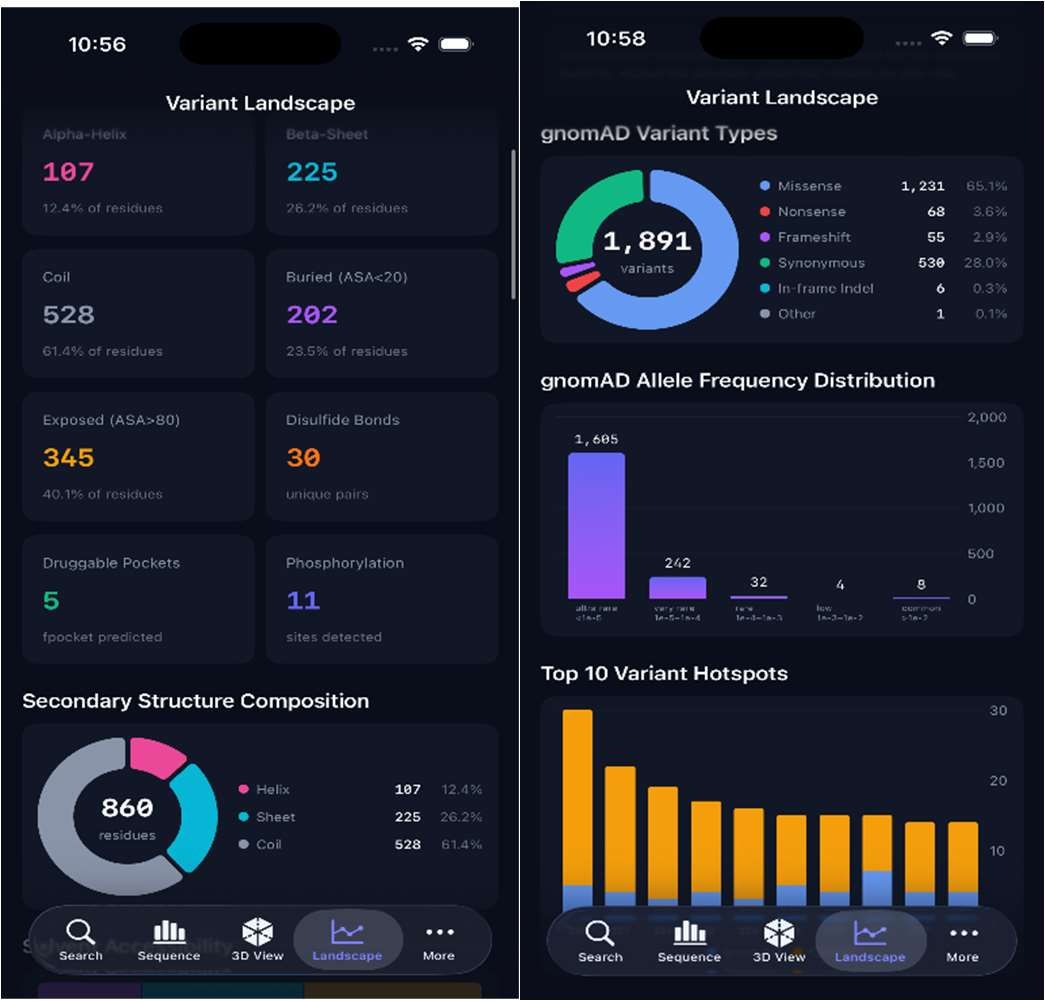}\\[0.4em]
{\footnotesize\sffamily\textbf{a}\quad Summary cards and gnomAD variant summary.}\\[0.7em]
\includegraphics[width=0.60\linewidth]{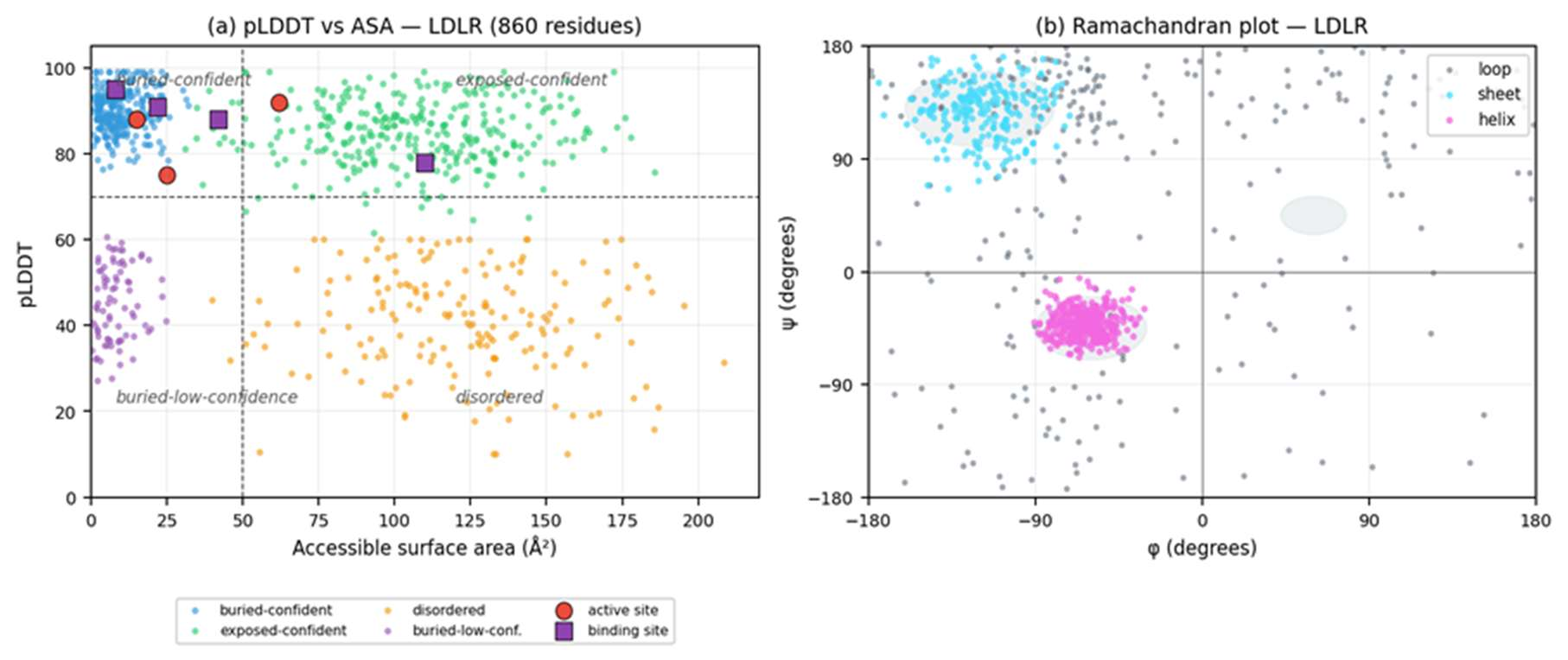}\\[0.25em]
{\footnotesize\sffamily\textbf{b}\quad pLDDT ASA quadrant scatter and Ramachandran plot.}
\caption{\textbf{Variant landscape dashboard for LDLR.} (\textbf{a})~Summary cards
report secondary structure composition, buried/exposed counts, disulfide bonds,
predicted druggable pockets, and post translational modification sites; the gnomAD
pane summarises protein consequence by type, the allele frequency distribution,
and the top variant hotspots. (\textbf{b})~The pLDDT ASA quadrant scatter labels
four biophysically meaningful populations and overlays active  and binding site
residues as enlarged markers, while the Ramachandran plot of residue
$(\phi,\psi)$ angles is coloured by three state secondary structure with canonical
allowed regions shaded.}
\label{fig:landscape}
\end{figure}

The Landscape module presents an analytical dashboard with no prior native mobile
counterpart (Fig.~\ref{fig:landscape}). Summary cards report helix/sheet/coil counts
and percentages, buried versus exposed positions (ASA below
\SI{20}{\angstrom\squared} versus above \SI{80}{\angstrom\squared}), unique disulfide
pairs, predicted pockets, and phosphorylation sites. A pLDDT versus ASA scatter
partitions the plane into four biophysically meaningful quadrants
(buried confident, exposed confident, buried low confidence, disordered) and overlays
active  and binding site residues as enlarged markers, so the intrinsically
disordered population (low pLDDT, high ASA) and functionally important outliers
become visible at once. An interactive Ramachandran panel places each residue in the
$(\phi,\psi)$ plane, bounded by $\pm 180^{\circ}$ and coloured by secondary structure
class; residues outside the allowed regions give a coarse stereochemical sanity check
on the AlphaFold prediction. Where a gnomAD summary is available, a companion pane
breaks down consequence by type and shows the allele frequency distribution and top
hotspots.

\subsection{Isoform reconstruction and druggable pocket exploration}
\begin{figure}[!tb]
\centering
\includegraphics[width=0.74\linewidth]{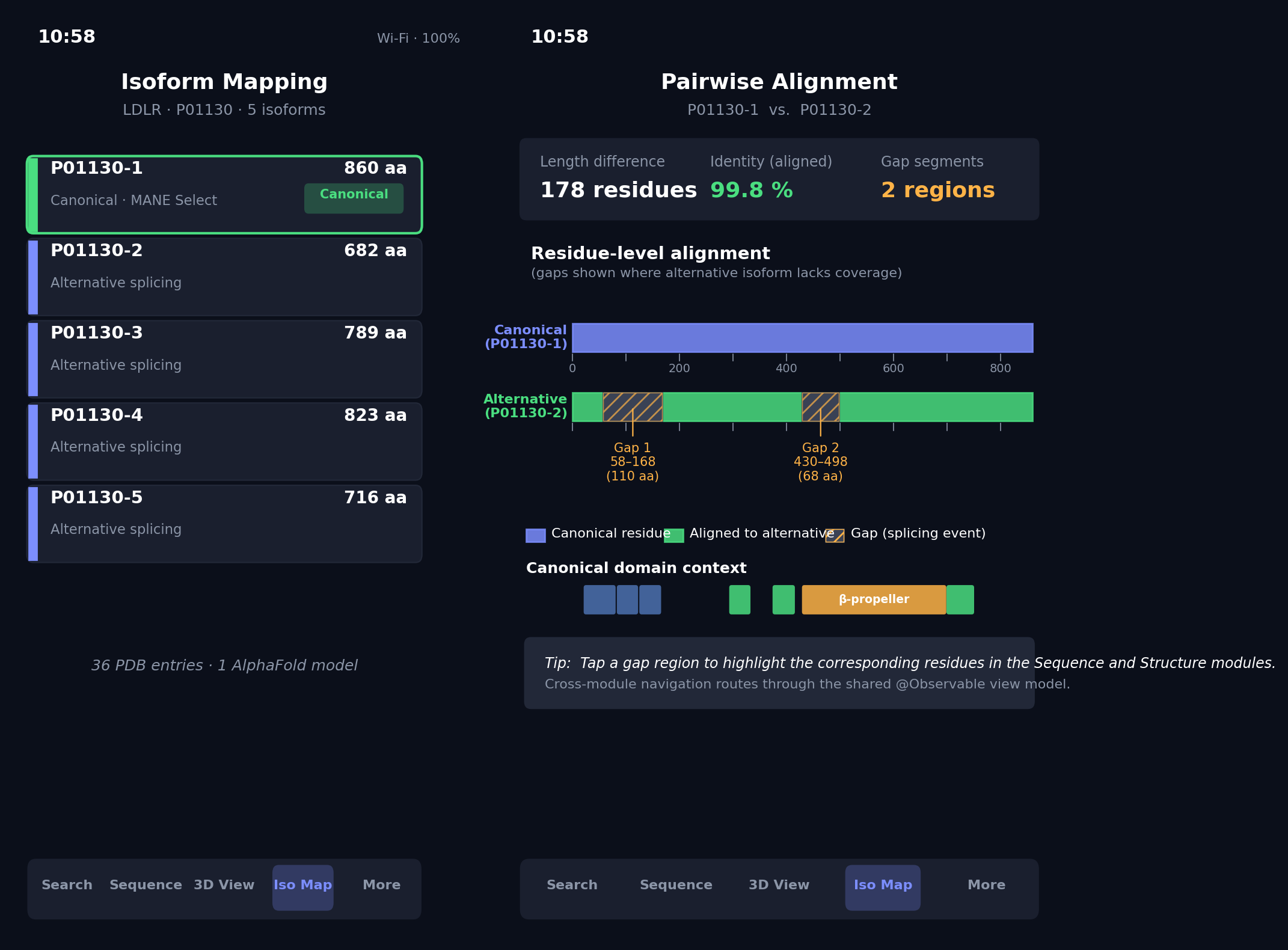}
\caption{\textbf{Pairwise isoform alignment reconstructed on device.} Left: the
Isoform module lists the five LDLR isoforms returned by the mapping endpoint and
flags the canonical, MANE Select entry. Right: the pairwise viewer renders the
residue level correspondence between the canonical (P01130 1) and alternative
(P01130 2) isoforms, with hatched segments marking the two splice associated gaps
that account for the 178 residue length difference; a canonical domain context bar
places the gaps in their structural neighbourhood.}
\label{fig:isoform}
\end{figure}

The Isoform module displays one card per isoform returned by the identifier mapping
endpoint, flags the canonical and MANE Select entries, and surfaces the
comma separated PDB chip list verbatim (Fig.~\ref{fig:isoform}). Because the portal
does not serve pairwise alignments to programmatic clients, a pairwise view
reconstructs the canonical versus alternative correspondence on device
(Section~\ref{sec:cs1}) and renders two parallel tracks with gaps as hatched regions.
For LDLR it reproduces the 178 residue difference between canonical P01130 1 (860
residues) and alternative P01130 2 (682 residues), localising it to two
splice associated gaps and reporting 99.8\% identity over aligned positions. This
context is critical: a residue numbering meaningful for one isoform may not exist in
another, and the canonical or MANE Select transcript may differ from the one
expressed in the tissue of interest.

\begin{figure}[!tb]
\centering
\includegraphics[width=0.74\linewidth]{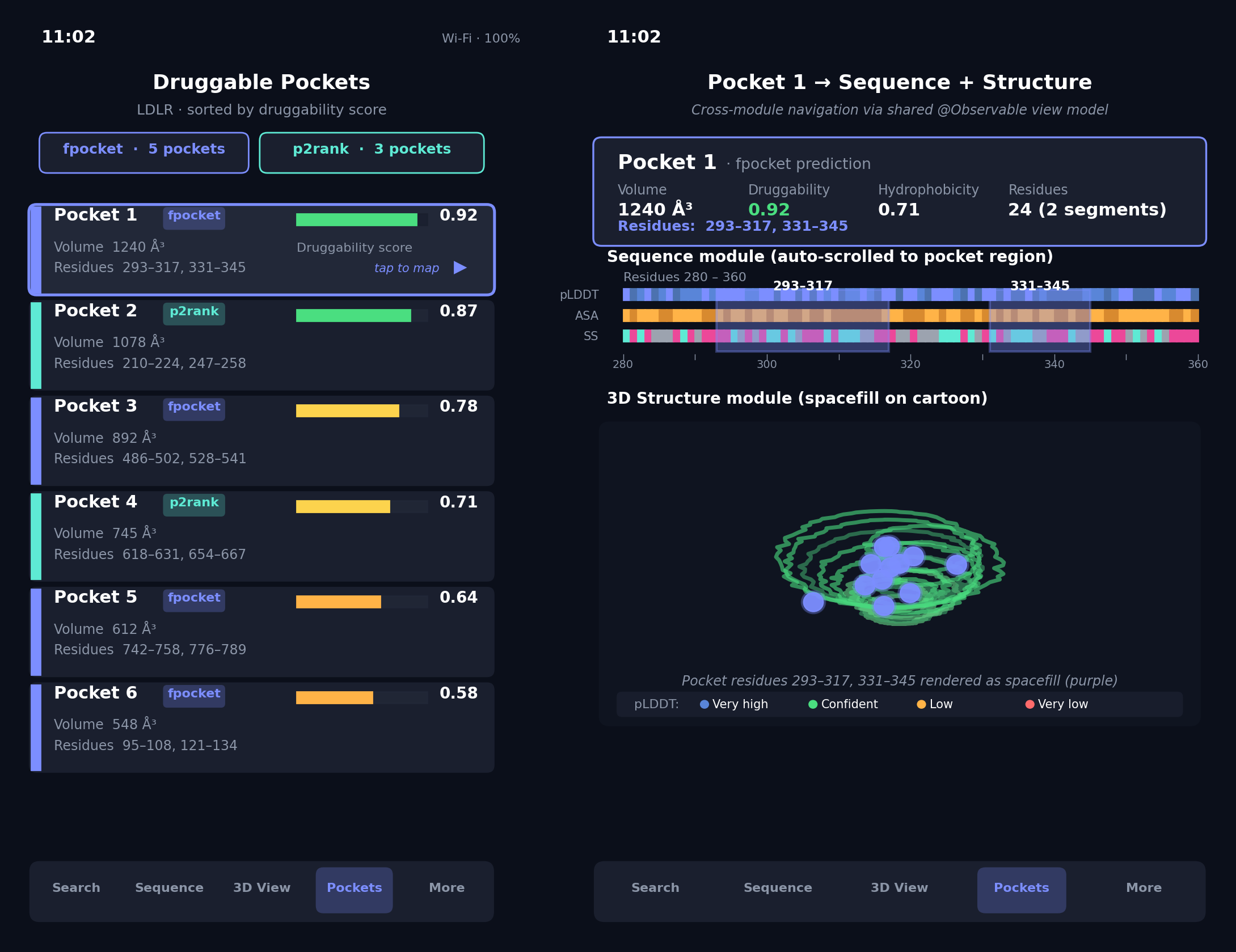}
\caption{\textbf{Druggable pocket explorer.} Left: fpocket and p2rank predictions
parsed from the seventy one column feature table, deduplicated and sorted by
druggability score; each card lists predictor source, pocket volume, score, and
constituent residues. Right: tapping the highest ranked pocket writes the residue
range to the shared view model, auto scrolling the Sequence module to the pocket
region (translucent overlay over the pLDDT, ASA, and secondary structure tracks)
and re issuing a Mol* selection that renders the pocket residues as spacefill on
the cartoon backbone. A predictor toggle switches between fpocket and p2rank.}
\label{fig:pockets}
\end{figure}

The Pockets module is the drug discovery oriented view (Fig.~\ref{fig:pockets}). The
portal stores fpocket and p2rank predictions\citep{LeGuilloux2009,Krivak2018} in the
same table, each pocket described by a free text cell such as ``Pocket 4: Mean pLDDT:
85.58, Volume: 477.18\,\AA$^3$, Druggability score: 0.05''. The module parses these
into structured records, deduplicates by pocket number, and presents one card per
pocket sorted by druggability score, showing predictor, volume, score as a gradient
bar, and constituent residues. Two actions (``Show on sequence'' and ``Show on 3D'')
write the residue range to the shared view model, scroll and highlight the Sequence
module, and re issue a Mol* selection rendering the pocket as spacefill against the
cartoon backbone replacing the desktop habit of many browser tabs with a single
gesture and no additional network requests.

\subsection{Engineering the desktop to mobile transition: three case studies}
\label{sec:casestudies}
\begin{figure*}[!t]
\centering
\includegraphics[width=0.94\linewidth]{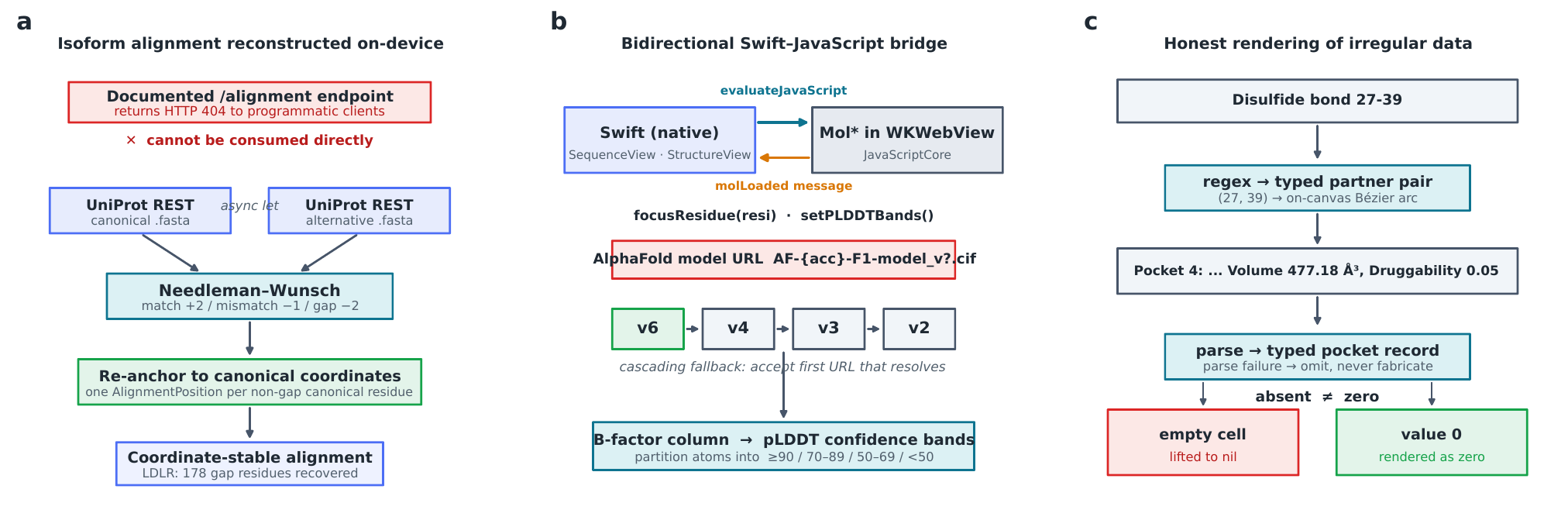}
\caption{\textbf{Three engineering challenges in bringing an aggregating web
service to the phone.} (\textbf{a})~The documented isoform alignment route returns
HTTP~404 to programmatic clients, so the client fetches canonical and alternative
FASTA sequences from UniProt, computes a Needleman wunsch alignment on device,
and re anchors it to canonical coordinates, recovering the 178 LDLR gap residues
without destabilising downstream indexing. (\textbf{b})~A bidirectional
Swift JavaScript bridge drives the Mol* viewer through \texttt{evaluateJavaScript}
and reports load completion back through a \texttt{molLoaded} message; a cascading
version fallback loader tries the AlphaFold model file versions in order and
accepts the first that resolves, and per residue pLDDT is recovered from the
mmCIF B factor column to filter confidence bands. (\textbf{c})~Semi structured
free text cells are parsed into typed records with parse failures omitted rather
than fabricated, and empty cells are lifted to \texttt{nil} so that an absent
annotation is never rendered as a value of zero.}
\label{fig:casestudies}
\end{figure*}

Three obstacles encountered during development were decisive enough to reshape the
architecture, and we present them as worked case studies because together they
characterise what it takes to consume an aggregating scientific web service
faithfully from a phone (Fig.~\ref{fig:casestudies}).

\subsubsection{Reconstructing an absent isoform alignment endpoint}
\label{sec:cs1}
The portal documents an isoform alignment endpoint and displays an alignment in its
desktop interface, yet the route returns HTTP~404 to programmatic clients. Inspecting
the portal's own network behaviour revealed why: the desktop alignment is computed in
the browser, which fetches the two isoform FASTA sequences from UniProt and aligns
them locally. A client that merely called the documented endpoint would render
nothing, and the fix could not be cosmetic, because the alignment feeds tracks in the
Isoform and Landscape modules that assume canonical numbering as a stable anchor. We
therefore mirrored the hidden behaviour on device: \texttt{fetchIsoformAlignment}
issues two concurrent requests for the canonical and alternative FASTA sequences,
strips headers, and passes them to a from scratch Needleman  Wunsch aligner\citep{Needleman1970}
(match $+2$, mismatch $ 1$, gap $ 2$) of roughly sixty lines of dependency free Swift.
A subtler step secures coordinate stability: because a global alignment inserts gaps
on both sequences, the projection routine advances independent canonical and
alternative counters and emits exactly one position per non gap canonical residue,
dropping pure canonical gap rows, so an insertion in the alternative isoform can never
shift the canonical index of downstream residues. The result is the alignment the API
does not serve, produced entirely on device, without destabilising the shared
residue coordinate system (Fig.~\ref{fig:isoform}).

\subsubsection{A bidirectional bridge across the WebKit boundary}
\label{sec:cs2}
Three dimensional structures are rendered by a JavaScript viewer in a
\texttt{WKWebView} while the rest of the app is native Swift, and three difficulties
followed. First, AlphaFold serves model files at version stamped URLs
(\texttt{AF \{accession\} F1 model\_v6.cif}), and the version advanced mid project, so
a hard coded version failed silently for re versioned entries. Second, the valuable
interaction tapping a residue and seeing it in 3D requires selection state to
travel from Swift into JavaScript and back. Third, per residue pLDDT is not a separate
field but is encoded in the mmCIF B factor column. The loader addresses the first by
trying \texttt{[v6, v4, v3, v2]} in order and accepting the first URL that responds,
with the request built from the base accession (any isoform suffix stripped, since the
structure service does not recognise suffixed identifiers). Selection propagates
through a minimal bidirectional bridge: a tap in \texttt{SequenceView} publishes the
residue; \texttt{StructureView} observes it and calls \texttt{focusResidue} through
\texttt{evaluateJavaScript}, zooming and marking it, while load completion returns
through a \texttt{WKScriptMessageHandler} named \texttt{molLoaded}. Confidence band
filtering (\texttt{setPLDDTBands}) reads each atom's B factor, partitions residues into
the four AlphaFold classes, and restyles each independently, so a confidence dip seen
in the sequence track is brought into spatial focus with one tap, tolerating both the
evolving versioning and the suffix convention without per entry intervention.

\subsubsection{Honest rendering of semi structured and missing data}
\label{sec:cs3}
Two data quality issues threatened misleading figures. First, several of the
portal's richest annotations arrive as free text embedded in tabular cells: disulfide
bonds as fragments like ``Disulfide bond 27 39'', and pockets as descriptors carrying
mean pLDDT, volume, and druggability score. Second, experimental PDB structures are
typically solved only for the canonical isoform, so a coverage chart plotting only
experimental structures would show empty bars for every alternative isoform, falsely
implying no structural model exists. The application could thus fail in two opposite
directions: crashing on irregular cells, or more dangerously silently rendering a
confident looking figure that overstates the data. We parse the free text cells with
targeted regular expressions into typed records (disulfide pairs feeding the arc
renderer, pocket descriptors feeding the Pockets module), omitting rather than
inventing on parse failure. The header (carrying the \AA$^2$ sign, asterisks, and
parenthesised qualifiers) is preserved verbatim so exact match keys stay valid, and
empty cells are lifted to \texttt{nil} so a missing annotation is never rendered as
zero. For coverage, each isoform is given a predicted model segment with a footnote
recording the predicted versus experimental asymmetry rather than concealing it. The
unifying principle across all three case studies is to \emph{localise each concession
to external reality behind a single typed boundary} (an aligner, a bridge function, a
parser) so the rest of the system operates on clean, typed data, and to fail visibly
rather than silently when a conversion is impossible.

\subsection{Quantitative performance evaluation}
\begin{figure}[!tb]
\centering
\includegraphics[width=0.76\linewidth]{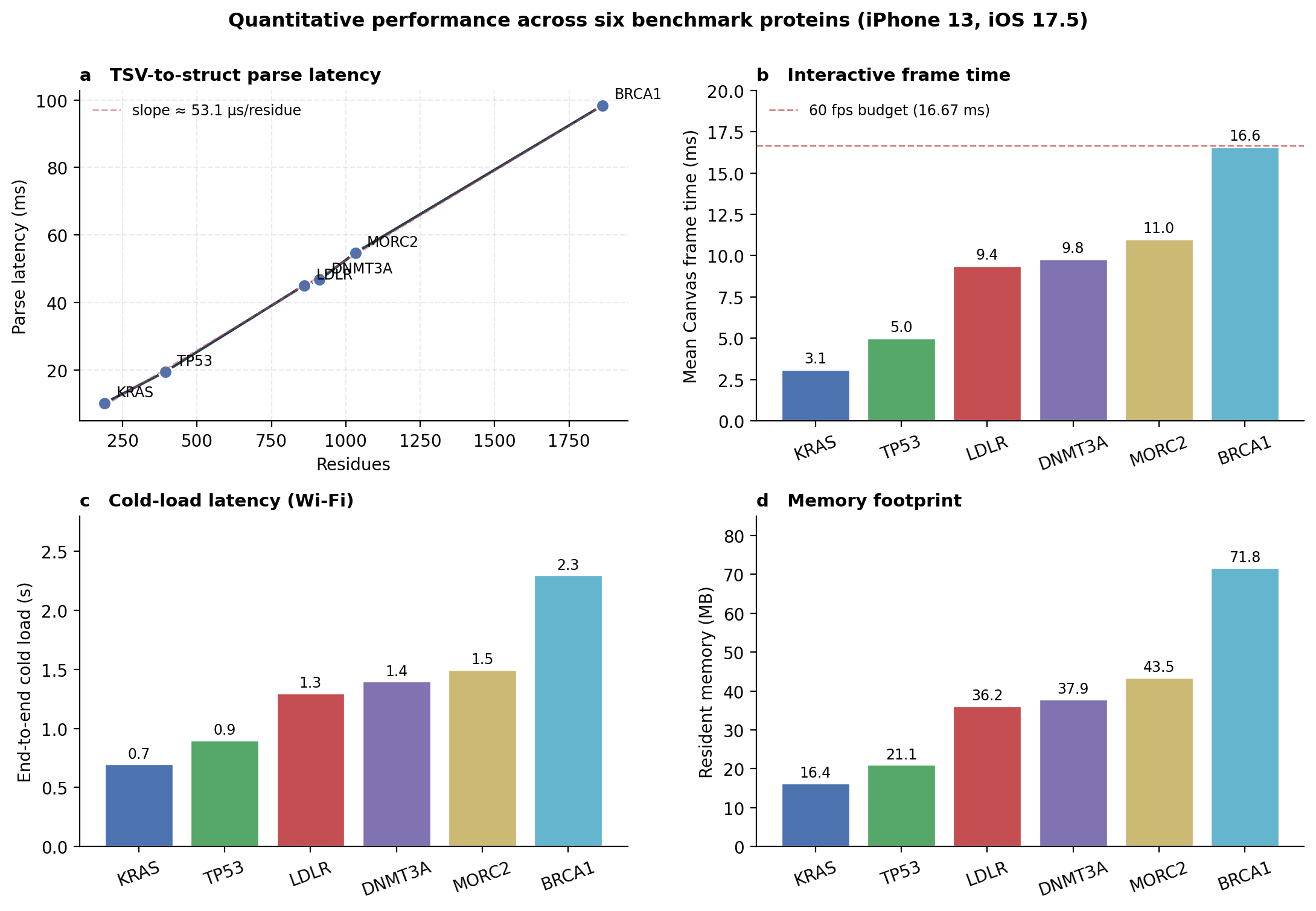}
\caption{\textbf{Quantitative performance across six benchmark proteins.}
(\textbf{a})~Parse latency scales near linearly with residue count at
$\sim$\SI{53}{\micro\second} per residue. (\textbf{b})~Mean \texttt{Canvas} frame
time during interactive zoom grows sub linearly because the visible viewport is
bounded; even BRCA1 (1{,}863 residues) sustains 16.6\,ms per frame, within the
60\,fps budget (dashed line). (\textbf{c})~End to end cold load time over Wi Fi
remains under 2.5\,s for all benchmarks. (\textbf{d})~Resident memory stays within
iOS foreground thresholds. iPhone~13, iOS~17.5, Release build, 200\,Mbps Wi Fi.}
\label{fig:perf}
\end{figure}

We benchmarked G2P Explorer across six proteins ranging from 189 to 1{,}863
residues on an iPhone~13 (Apple A15 Bionic, 4\,GB RAM, iOS~17.5) over a
200\,Mbps Wi Fi network in Release configuration
(Table~\ref{tab:perf}, Fig.~\ref{fig:perf}). The application was force quit and
cold launched between proteins; interactive frame timing was recorded with
Instruments Time Profiler over thirty seconds of continuous pinch zoom and scroll
with the Sequence module front most.

\begin{table}[!tb]
\centering
\caption{\textbf{Benchmark proteins and quantitative performance.} The six proteins
span the residue length distribution of human protein coding genes and a range of
disease contexts (KRAS, signalling GTPase and cancer driver; TP53, multi domain
tumour suppressor; LDLR, familial hypercholesterolaemia and primary validation
target; DNMT3A, DNA methyltransferase; MORC2, Charcot Marie Tooth type~2Z;
BRCA1, large protein with an intrinsically disordered central region). Cold load
is wall clock from preset tap to first rendered frame over Wi Fi; parse latency is
the tab separated to struct conversion time; frame time is the mean \texttt{Canvas}
draw time during thirty seconds of continuous pinch zoom and scroll; memory is the
resident set size with the Sequence module front most and the Mol* viewer loaded.
iPhone~13, iOS~17.5, Release build, 200\,Mbps Wi Fi.}
\label{tab:perf}
\small
\begin{tabularx}{\linewidth}{l l R R R R R R}
\toprule
Protein & UniProt & Residues & Payload (KB) & Cold load (s) & Parse (ms) & Frame (ms) & Memory (MB) \\
\midrule
KRAS    & P01116 & 189     & 62  & 0.7 & 10.2 & 3.1  & 16.4 \\
TP53    & P04637 & 393     & 138 & 0.9 & 19.5 & 5.0  & 21.1 \\
LDLR    & P01130 & 860     & 347 & 1.3 & 45.1 & 9.4  & 36.2 \\
DNMT3A  & Q9Y6K1 & 912     & 362 & 1.4 & 46.8 & 9.8  & 37.9 \\
MORC2   & Q9Y6X9 & 1{,}032 & 421 & 1.5 & 54.7 & 11.0 & 43.5 \\
BRCA1   & P38398 & 1{,}863 & 748 & 2.3 & 98.4 & 16.6 & 71.8 \\
\bottomrule
\end{tabularx}
\end{table}

Parse latency scales near linearly with residue count at approximately
\SI{53}{\micro\second} per residue, from 10.2\,ms for KRAS to 98.4\,ms for BRCA1.
End to end cold load time ranges from 0.7\,s to 2.3\,s and is dominated by the
200 900\,ms network stage rather than by on device work. Mean \texttt{Canvas}
frame time during interactive zoom grows sub linearly with protein length, because
the visible viewport is bounded; even for BRCA1 the renderer sustains 16.6\,ms per
frame, within the 60\,fps budget. Resident memory ranges from 16.4\,MB to 71.8\,MB,
well within iOS foreground thresholds. Across the full benchmark, the framework
therefore sustains interactive frame rates and modest memory without resorting to
residue binning or paging.

\subsection{Scope comparison and application to disease proteins}
We compared the scope of G2P Explorer with five widely used visualization tools
(Table~\ref{tab:compare}). PyMOL\citep{Yuan2017} and ChimeraX\citep{Pettersen2021}
remain the strongest desktop molecular environments, with scripting and rendering no
mobile framework approaches; Mol*\citep{Sehnal2021} is the de facto web viewer for
AlphaFold and PDB content; ProtVista\citep{Watkins2017} is the reference
multi track sequence library in the browser; and IGV\citep{Thorvaldsdottir2013} is
the standard genome browser for variant data. None was designed as a touch first
mobile framework, and none combines residue level G2P annotations, AlphaFold
rendering, on device isoform alignment, and pocket exploration in one environment.
The comparison clarifies the design space rather than claiming superiority: for deep
structural analysis on a workstation, PyMOL, ChimeraX, and desktop Mol* remain the
right choices.

\begin{table}[!tb]
\centering
\caption{\textbf{Scope comparison with established sequence and structural
visualization tools.} The comparison clarifies the niche occupied by G2P Explorer
rather than positioning it as a replacement for the deeper desktop viewers.
Capabilities listed are typical of each tool's primary use case; plug ins or
extensions may extend any individual tool beyond what is shown.}
\label{tab:compare}
\small
\begin{tabularx}{\linewidth}{X c c c c c >{\columncolor{accent!8}}c}
\toprule
Capability & PyMOL & ChimeraX & Mol* & ProtVista & IGV & \textbf{G2P Explorer} \\
\midrule
Primary platform        & Desktop & Desktop & Web & Web & Desktop & \textbf{iOS native} \\
Touch first UI          & No  & No  & Partial & Partial & No  & \textbf{Yes} \\
Multi track sequence    & Limited & Limited & Limited & Yes & Yes & \textbf{Yes (6 tracks)} \\
3D AlphaFold viewing    & Yes & Yes & Yes & No  & No  & \textbf{Yes (Mol* bridge)} \\
Residue annotation      & Manual & Manual & Limited & Yes & Limited & \textbf{Yes (71 cols)} \\
Isoform alignment       & No  & No  & No  & Partial & No  & \textbf{Yes (on device)} \\
Druggable pockets       & Plug in & Plug in & No  & No  & No  & \textbf{Yes (fpocket/p2rank)} \\
Ramachandran plot       & Plug in & Built in & No  & No  & No  & \textbf{Yes (interactive)} \\
Offline mode            & Yes & Yes & No  & No  & Yes & \textbf{Cache + sample} \\
\bottomrule
\end{tabularx}
\end{table}

\begin{figure}[!tb]
\centering
\includegraphics[width=0.64\linewidth]{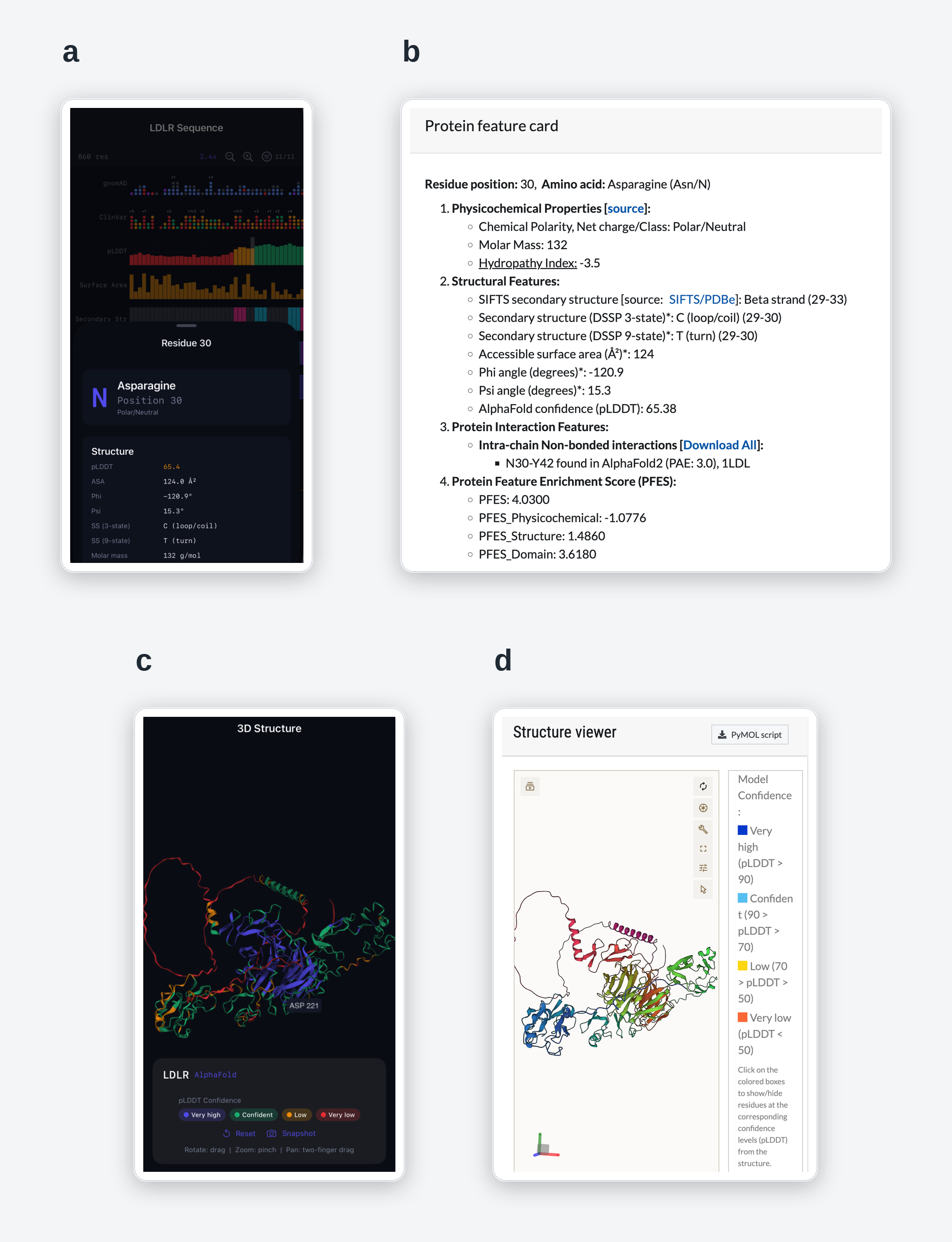}
\caption{\textbf{Head to head interface comparison of G2P Explorer (iOS) and the
G2P web portal for LDLR residue~30.} (\textbf{a})~The residue detail card in G2P
Explorer presents amino acid identity, physicochemical class, and per residue
structural metrics (pLDDT, ASA, $\phi/\psi$, three  and nine state secondary
structure) as a compact, scannable sheet. (\textbf{b})~The equivalent G2P web
``Protein feature card'' lists the same fields as a dense text block.
(\textbf{c})~The mobile Structure module renders the AlphaFold model with pLDDT
colouring and a minimal control surface suited to a phone. (\textbf{d})~The desktop
structure viewer retains the full control panel and legend. Both clients read
identical values from the public G2P data and differ only in how legibly those
values reach a small screen.}
\label{fig:appcompare}
\end{figure}

\begin{figure}[!tb]
\centering
\includegraphics[width=0.82\linewidth]{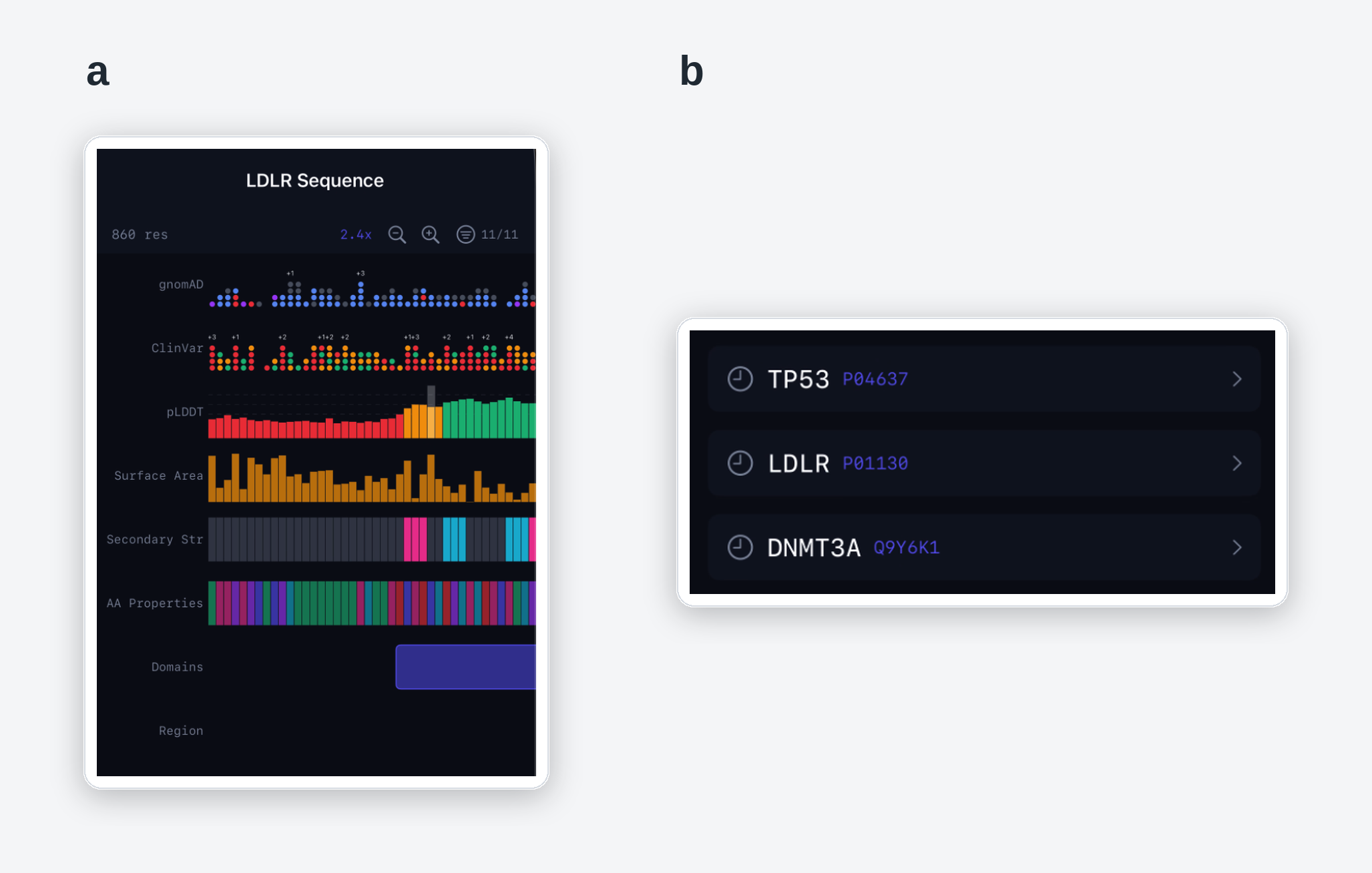}
\caption{\textbf{Mobile native conveniences absent from the desktop portal.}
(\textbf{a})~The multi track sequence renderer in landscape orientation under
pinch zoom (shown at 2.4$\times$); all 860 LDLR residues remain addressable across
six tracks with no binning or paging. (\textbf{b})~The Search module's
recent protein list: entries marked with a clock icon were fetched earlier and are
served from the on device cache, so they reopen instantly and stay available with
no network connection.}
\label{fig:mobile}
\end{figure}

Table~\ref{tab:compare} situates G2P Explorer among general purpose viewers.
Because the framework is a mobile re presentation of the G2P web
portal\citep{Kwon2024} in particular, we also compared the two directly, feature
by feature (Table~\ref{tab:webcompare}, Figs.~\ref{fig:appcompare} and
\ref{fig:mobile}). Both clients expose the same residue level data, yet differ in
how legibly it reaches a small touch screen. The portal renders a residue as a
dense text block (Fig.~\ref{fig:appcompare}b), whereas G2P Explorer presents the
same fields as a structured, scannable card (Fig.~\ref{fig:appcompare}a); the
portal keeps the full desktop control surface on the structure viewer
(Fig.~\ref{fig:appcompare}d), while the mobile viewer reduces it to the essentials
for residue reading (Fig.~\ref{fig:appcompare}c). Three conveniences have no
counterpart in the browser tool: an on device cache that reopens a previously
viewed protein instantly and offline, one gesture pinch zoom across all residues,
and a landscape multi track layout (Fig.~\ref{fig:mobile}).

\begin{table}[!tb]
\centering
\caption{\textbf{Direct comparison of G2P Explorer (iOS) with the upstream G2P web
portal.} Both clients consume the same public G2P data; the differences concern
presentation and mobile ergonomics rather than the underlying annotations.}
\label{tab:webcompare}
\small
\begin{tabularx}{\linewidth}{X l >{\columncolor{accent!8}}l}
\toprule
Capability & G2P web portal & \textbf{G2P Explorer (iOS)} \\
\midrule
Offline reopen / caching     & No               & \textbf{Yes} \\
Scalable pinch zoom          & Not readily      & \textbf{One gesture} \\
Residue detail layout        & Dense text block & \textbf{Structured card} \\
Landscape multi track view   & No               & \textbf{Yes} \\
Isoform alignment comparison & No               & \textbf{Yes (on device)} \\
Isoform mapping              & Manual           & \textbf{Automatic} \\
Glass morphism interface     & No               & \textbf{Yes} \\
Dark theme                   & No               & \textbf{Yes} \\
\bottomrule
\end{tabularx}
\end{table}

Exercising all six modules across the benchmark proteins confirmed correctness and
the integrated workflow. For LDLR, the merged domain track recovers the alternating
LDL receptor class A repeats, EGF like repeats, and the C terminal $\beta$ propeller
in agreement with UniProt; the disulfide lane exposes the dense class A disulfide
network as arcs; and the Isoform module lists five isoforms with a thirty six entry
PDB list and the correct 178 residue gap. BRCA1, at nearly twice the residue count,
stresses the renderer without frame rate regression and immediately exposes its long
disordered central region as a dense population in the disordered quadrant, while the
C terminal BRCT domain forms a confident cluster. TP53 exercises small multi domain
proteins: the DNA binding domain is delineated and the Ramachandran panel is confined
to the canonical $\alpha$ and $\beta$ regions. MORC2 and DNMT3A, the biological case
studies of the original portal manuscript\citep{Kwon2024}, load through the identical
code path; the Sequence module reveals the GHKL ATPase domain in MORC2 and the
methyltransferase domain in DNMT3A, and the Pockets module surfaces several cavities
for DNMT3A consistent with the tractability of the methyltransferase fold. No
biological discovery is claimed; the workflow shows how the integrated mobile loop
supports rapid hypothesis generation on one device with no browser tabs or manual
cross referencing.

\subsection{Validation of the engineering solutions}
Beyond module level functionality, the three engineering solutions of
Section~\ref{sec:casestudies} were each checked against an observable ground truth.
The isoform reconstruction (Section~\ref{sec:cs1}) was validated against LDLR, whose
canonical and second isoforms differ by a well characterised splice event: the
on device alignment reproduces the expected 178 gap residues at the splice boundaries
and matches the bundled reference. The version fallback loader (Section~\ref{sec:cs2})
was exercised across the database's own model file transition; entries returning 404
under the previously hard coded version resolved once the cascading loader was
introduced, and no preset protein failed to load afterwards. The honest rendering
measures (Section~\ref{sec:cs3}) were checked by confirming that alternative isoforms,
which carry no experimental structures, render a predicted model segment with the
asymmetry footnote rather than an empty bar, and that malformed or empty cells produce
an omitted glyph rather than a crash or a spurious zero.

\section{Discussion}

G2P Explorer demonstrates that the integrated genomics to proteomics workflow of
the G2P portal can be redesigned for touch hardware without sacrificing
residue level fidelity. It is not a port of the desktop portal but a re imagining of
the same data products through a touch first interaction grammar, and that
distinction matters: mobile native scientific visualization opens usage contexts
desktop tools cannot reach the bedside, the conference hallway, the lecture room,
the field site each a moment of interpretation in which quick reference to a
variant's structural environment could change a downstream decision. The on canvas
disulfide arcs, the quadrant labelled pLDDT ASA scatter, and the interactive
Ramachandran panel are, to our knowledge, the first mobile native realisations of
these idioms, and cross module residue linking collapses a multi tab desktop
workflow into a single gesture. Visualization here is not the last step of an
analysis but an active driver of hypothesis generation: the buried low confidence
quadrant is a population that calls for explanation, and the Ramachandran panel is a
stereochemical sanity check rather than decoration.

The framework is explicitly complementary to, not a replacement for, the G2P portal
and the established desktop and web viewers. The portal remains the authoritative
source of the underlying data\citep{Kwon2024}; our framework re presents those
products in a mobile native form, inheriting the portal's data licensing and update
cadence by relying on its public API rather than a proprietary backend. The three
case studies suggest a design principle that generalises. Each obstacle we
encountered a missing API endpoint, a shifting remote versioning scheme, a
domain specific encoding buried in a structure file, semi structured free text
cells, isoform asymmetric coverage is a concession to external reality the
application cannot eliminate but must absorb. The architecture survives these
concessions because each is localised behind a single typed boundary, so the rest of
the system never sees the irregularity. The transferable lesson for anyone building
a native client over an aggregating scientific web service: do not let upstream
irregularities propagate inward; convert them, once, at the boundary, into clean
typed values, and fail visibly rather than silently when conversion is impossible.

Several limitations should be stated plainly. The framework currently targets iOS;
an Android port would require a separate implementation, most naturally with Jetpack
Compose Canvas. It depends on the public G2P REST API, so service outages or schema
changes will degrade live functionality, although the bundled sample and the
fault tolerant ingestion partially mitigate this. The embedded Mol* viewer runs
inside a \texttt{WKWebView} subject to JavaScriptCore ceilings rather than native
Metal acceleration, so very large assemblies may render more slowly than in a
desktop Mol* session, and 3D rendering still requires connectivity because both the
library and the AlphaFold model are fetched at run time, even though feature and
isoform data are now served from the on device cache. The framework presents static
AlphaFold models and does not represent conformational dynamics, ensemble
predictions, or pathogenicity scores such as AlphaMissense\citep{Cheng2023}; it
surfaces evidence rather than a verdict. Finally, the evaluation is quantitative for
performance but developer conducted for usability and uses six proteins rather than
a proteome sweep, so the usability observations are formative; the annotations
surfaced are not clinical diagnoses and should not be used as such.

Several improvements follow directly. The current cache already covers feature and
isoform data; bundling the Mol* library and a few pre fetched AlphaFold models, with
on disk caching of downloaded mmCIF files, would extend the same instant reload
benefit to the structural view. Generalising the isoform reconstruction from
canonical versus one alternative to proteins with many isoforms, and replacing the
fixed preset grid with validated arbitrary accession entry, would broaden coverage.
Integrating variant effect predictors (AlphaMissense\citep{Cheng2023}, REVEL, CADD)
as additional tracks, and structure based search through
Foldseek\citep{VanKempen2024} or an on device pathogenicity model, would convert the
tool from a reference instrument into an analytical one. Finally, a controlled
usability study with clinicians, educators, and researchers would turn these
formative observations into summative evidence. The Swift code base, parser,
on device aligner, and Mol* bridge are intended for release as a reproducible
reference implementation for future mobile bioinformatics platforms.

\section{Methods}
This section describes the methodology adopted for the development and implementation of the proposed framework. It outlines the data sources, system architecture, data processing pipeline, and core implementation strategies employed throughout the study.

\subsection{Data sources and API endpoints}
G2P Explorer consumes tab-separated REST endpoints exposed by the G2P
portal\citep{Kwon2024}: a protein features endpoint returning per-residue
annotations and an identifier mapping endpoint returning
gene transcript protein isoform structure mappings; both return tab-separated
tables rather than JSON, so the client implements its own parser (below).
Three-dimensional rendering uses AlphaFold mmCIF files retrieved directly from the
EBI AlphaFold service\citep{Varadi2024} using the UniProt accession with any
isoform suffix stripped, and experimental structures can optionally be loaded by
accession from the RCSB PDB\citep{Berman2000}. The identifier mapping endpoint
returns canonical transcripts flagged with an asterisk and MANE Select transcripts
with a label\citep{Morales2022}; these flags are part of the cell value, not a
separate column, and the parser preserves them verbatim. The pairwise
isoform alignment route documented by the portal returns HTTP~404 to programmatic
clients, so the client reconstructs the alignment on device from UniProt FASTA
sequences\citep{UniProt2023} (Section~\ref{sec:cs1}).

\subsection{Tab separated parsing and data model}
The \texttt{TSVParser} splits rows on newline characters, treats the first
non-blank row as the header, and splits both header and data rows on tab
characters. The header is preserved verbatim, including non-ASCII characters such
as the \AA$^2$ symbol, asterisks denoting the canonical isoform, and parenthesised
source qualifiers, so that exact match lookup is possible at the model layer. Empty
cells are lifted to Swift \texttt{Optional} values through two helpers
(\texttt{optionalString}, \texttt{optionalDouble}), preserving the semantic
distinction between ``no annotation'' and ``annotation value is zero'' that a naive
cast to \texttt{Double} would erase. Unknown columns do not raise an error; they
are stored verbatim in a \texttt{rawColumns} dictionary on the
\texttt{ProteinFeature} struct so that they remain available in the residue detail
sheet (Supplementary Fig.~1). The parser is content agnostic and is reused for
every tab-separated endpoint. Data models are pure Swift structs conforming to
\texttt{Identifiable} and \texttt{Hashable}; \texttt{IsoformMapping} parses the
comma-separated PDB column on construction, and \texttt{IsoformAlignment} exposes a
derived property that groups consecutive gap residues into closed integer ranges
suitable for \texttt{Canvas} rendering.

\subsection{Client side isoform alignment}
\texttt{G2PService.fetchIsoformAlignment} issues two concurrent requests
(\texttt{async let canonFasta} and \texttt{async let altFasta}) to the UniProt
REST service, strips the FASTA headers, and passes the sequences to a from-scratch
Needleman Wunsch global aligner\citep{Needleman1970}
(\texttt{SequenceAligner.needlemanWunsch}; match $+2$, mismatch $ 1$, gap $ 2$)
implemented in roughly sixty lines of dependency-free Swift. A projection routine
then advances independent canonical and alternative counters and emits exactly one
\texttt{AlignmentPosition} per non-gap canonical residue, dropping pure canonical gap
rows, so that an insertion in the alternative isoform cannot shift the canonical
index of downstream residues. For LDLR, this reproduces the 178 gap residues
between isoforms one and two at the splice boundaries and matches the bundled
offline reference.

\subsection{Mol* bridge and AlphaFold version fallback}
The Structure module hosts an embedded Mol* viewer\citep{Sehnal2021} inside a
\texttt{WKWebView}. A locally bundled HTML file imports Mol*~4.9.0 from a CDN and
exposes \texttt{loadProtein}, \texttt{resetView}, \texttt{focusResidue}, and
\texttt{setPLDDTBands} to the Swift layer; Swift drives the viewer through
\texttt{evaluateJavaScript} and subscribes to molLoaded messages through
the WKScriptMessageHandler protocol. The model loader requests
AF \{accession\} F1 model\_v\{6,4,3,2\}.cif in order and accepts the first
URL that resolves, constructing the request from the base accession with any
isoform suffix stripped. Confidence band filtering reads each atom's B factor,
partitions residues into the four AlphaFold confidence classes
($\geq$90, 70 89, 50 69, $<$50), and restyles each class independently
(Section~\ref{sec:cs2}). To keep the view legible on a small screen, the bridge
exposes only the controls needed for residue level reading and draws the model in
a reduced form, so the structural view stays simpler than the full desktop viewer
while still showing the cartoon model, the chosen coloring, and any residue placed
in focus.

\subsection{Sequence rendering and analytical panels}
The Sequence module draws six tracks in immediate mode through SwiftUI's Canvas, issuing one filled rectangle per residue per track via
context.fill(Path(rect)). Pinch zoom is implemented as a
.scaleEffect modifier on a horizontal ScrollView, so SwiftUI
applies a single transformation rather than relaying out residues on each zoom
step. Per residue text labels are reserved for tap-driven detail sheets and for
domain rectangles wider than sixty points. Expensive per-protein-derived
values (the dataset maximum ASA used for normalisation, alignment gap ranges, and
parsed pocket records) are computed once on load and cached on the view model
rather than recomputed inside the Canvas body. Disulfide bonds and pocket
descriptors are extracted from free text with regular expressions and rendered as
B\'ezier arcs and pocket cards respectively, with parse failures omitted. The
landscape panels render directly from the parsed columns: the pLDDT ASA scatter
from the pLDDT and ASA columns with quadrant thresholds at
\SI{50}{\angstrom\squared} ASA and 70 pLDDT, and the Ramachandran panel from the
$\phi$ and $\psi$ columns coloured by three-state secondary structure.

\subsection{iOS implementation and offline support}
The framework is implemented in Swift~5.9 with SwiftUI for iOS~16 and above, with no
third-party Swift packages. The six modules share a single ProteinViewModel
(@Observable); networking is concentrated in the G2PService actor,
which wraps a URLSession (sixty seconds per request, one hundred twenty seconds
per resource timeouts, and default URL cache) and exposes asynchronous methods returning
typed values with structured concurrency cancellation from the SwiftUI lifecycle.
For offline demonstration, three reference tab-separated files for LDLR are bundled
and read through the same parser as live responses, so online and offline paths
exercise identical structures. Beyond this sample, live results are cached on device:
on a successful fetch, the raw feature and mapping responses are written to the cache's
directory under a key derived from the gene symbol or accession, and a later request
decodes the saved response through the same parser, so a previously viewed protein
loads from disk and stays available offline. The cache holds the data used by the
four non-structural modules; the 3D model is still fetched at run time because the
Mol* library and AlphaFold coordinates are retrieved over the network. Dark mode is
enforced at the root because the visualizations were designed against a dark palette.
The interface was verified on an iPhone~SE and on an iPad in both orientations with
width-adaptive layouts.

\subsection{Evaluation protocol}
Six proteins were selected to span the residue length distribution of human
protein-coding genes: KRAS (189 residues), TP53 (393), LDLR (860), DNMT3A (912),
MORC2 (1{,}032), and BRCA1 (1{,}863). All measurements were performed on an
iPhone~13 (Apple A15 Bionic, 4\,GB RAM, iOS~17.5) connected to a 200\,Mbps Wi Fi
network, with the application built in Release configuration; the application was
force-quit and cold-launched between proteins. End-to-end cold load latency was
measured wall clock from preset tap to first rendered frame. Parse latency was
instrumented around the TSVParser entry and exit points using
ContinuousClock and reflects tab-separated to struct conversion only. Mean
frame time during interactive zoom was recorded with Instruments Time Profiler over
thirty seconds of continuous pinch zoom and scroll with the Sequence module
frontmost. Resident memory was reported by Instruments Allocations while the
Sequence module was foremost and the Mol* viewer was loaded.

\section*{Data availability}
\addcontentsline{toc}{section}{Data availability}
The upstream G2P REST API is publicly available at
\url{https://g2p.broadinstitute.org/} and documented at
\url{https://g2p.broadinstitute.org/api docs/}\citep{Kwon2024}. AlphaFold mmCIF
files are retrieved from the EBI AlphaFold service
(\url{https://alphafold.ebi.ac.uk/})\citep{Varadi2024}; isoform FASTA sequences are
retrieved from the UniProt REST service
(\url{https://rest.uniprot.org/})\citep{UniProt2023}. The application bundles a
reference data set for LDLR (UniProt P01130) for offline demonstration. No new
biological data were generated for this study.

\section*{Code availability}
\addcontentsline{toc}{section}{Code availability}
The G2P Explorer source code, including the tab separated parser, the on device
Needleman Wunsch aligner, the Mol* bridge, and the benchmark instrumentation, will
be deposited in a public repository upon acceptance and is available from the
corresponding author on reasonable request. The framework depends on no
third party Swift packages; the Mol* JavaScript bundle (v4.9.0) is loaded from a
CDN at run time.


\section*{Acknowledgements}
\addcontentsline{toc}{section}{Acknowledgements}
We thank the authors of the Genomics 2 Proteins portal for designing and
maintaining the public API and the underlying data integration that this work
consumes, and the AlphaFold team at Google DeepMind and EMBL EBI, the RCSB PDB,
UniProt, Ensembl, gnomAD, ClinVar, the developers of fpocket and p2rank, and the
maintainers of Mol*, whose data products and software are essential to the
framework described here.

\section*{Author contributions}
A.A.E. and M.A.H. designed and implemented the framework, conducted the evaluation, and prepared the figures. M.M.H. supervised the project, contributed to the design and analysis, and edited the manuscript. All authors reviewed and approved the final manuscript.

\section*{Competing interests}
The authors declare no competing interests.



\footnotesize
\bibliographystyle{naturemag}
\bibliography{references}
\normalsize

\end{document}